\documentstyle[12pt]{article}
\begin{document}
\newcommand{\beq}{\begin{equation}}
\newcommand{\eeq}{\end{equation}}
\newcommand{\bea}{\begin{eqnarray}}
\newcommand{\eea}{\end{eqnarray}}
\newcommand{\eps}{\varepsilon}
\newcommand{\Veff}%
{{\cal V}^{\mbox{\scriptsize p}}_{\mbox{\scriptsize eff}}}
\newcommand{\Afr}{A^{\mbox{\scriptsize fr}}}
\newcommand{\Bfr}{B^{\mbox{\scriptsize fr}}}
\newcommand{\Binf}{B^{\mbox{\scriptsize inf}}}
\newcommand{\Blp}{B^{\mbox{\scriptsize LPA}}}
\newcommand{\Fs}{\mbox{\scriptsize F}}
\newcommand{\inn}{\mbox{\scriptsize in}}
\newcommand{\exx}{\mbox{\scriptsize ex}}

\centerline{\large\bf Can weakly bound heavy nuclei with very large}
\centerline{\large\bf neutron excess exist?  }

\vskip 0.5 cm

\centerline{M.~Baldo$^{a}$, U.~Lombardo$^{b,c}$,
E.E.~Saperstein$^{d}$ and M.V.~Zverev$^{d}$}
\vskip 0.5 cm

\centerline{$^a$INFN, Sezione di Catania, 57 Corso Italia,
I-95129 Catania, Italy}
\centerline{$^b$INFN-LNS, 44 Via S.-Sofia, I-95123 Catania, Italy}
\centerline{$^c$Dipartimento di Fisica, 57 Corso Italia,
I-95129 Catania, Italy}
\centerline {$^{d}$ Kurchatov Institute, 123182, Moscow, Russia }
\vskip 0.5 cm

\centerline{Abstract}
\vskip 0.5 cm
A realistic model is suggested based on the quasiparticle
Lagrange version of the self-consistent Finite Fermi
Systems theory supplemented with the microscopically 
calculated surface parameters of the Landau-Migdal interaction
amplitude. The latter are expressed in terms of the off-shell
$T$-matrix of free $NN$-scattering and show a strong dependence on the
chemical potential of a nucleus under consideration in the
drip line vicinity. This effect could result in shifting
the neutron drip line position to very large values of the
neutron excess.

\newpage
\vskip 0.5 cm
Up to now, all predictions on the location of the nuclear drip 
line for heavy nuclei are based on phenomenological 
approaches. All of them operate
with parameters adjusted to the properties of stable nuclei.
This point seems to be questionable for nuclei far from
the $\beta$-stability valley.
In view of commonly recognized importance of pairing for the
drip line, the 
main efforts in this field were focused to study the
superfluidity effects at small values
of the chemical potential $\mu$ \cite{Dob1, Dob2, Fay1}.
Such problems were investigated as a correct account of continuum, 
comparison of exact Bogolyubov solutions versus
those within the BCS approximation, the surface pairing versus
the volume one, and so on.
In this letter we concentrate on the examination of 
the neutron average potential well at
small values of $\mu_n$. We
give arguments in favour of significant
variation, in the vicinity of the drip line,
of some parameters 
of the effective $NN$-interaction which generates this
potential well. It could result in shifting the position of the
drip line into the region of large 
(or, maybe, very large)  values of the neutron excess.

The reason for such an effect can be readily shown in
terms of the simplest version of the self-consistent (SC)
Finite Fermi Systems (FFS) theory \cite{AB1,AB2}
which is based on the simplified version of the self-consistency 
relation of Ref.~\cite{FKh}:
\beq
\frac {\partial U} {\partial { \bf r} }
= \int F({\bf r},{\bf r'})
\frac {\partial \rho} {\partial { \bf r'} }\,  d { \bf r'} , 
\label{SCR}
\eeq
where $F$ is the Landau-Migdal (LM) amplitude.
Obvious isotopic indices in Eq.~(\ref{SCR}) are omitted.
Let us for a while limit ourselves to the zero-range components 
of $F$, which in the standard FFS theory notation has the form:
\beq
F({\bf r},{\bf r'}) = C_0 \left[ f_0({\bf r}) +f'_0({\bf r}) 
{\bf \tau_1} {\bf \tau_2}+
( g_0({\bf r}) + g'_o({\bf r}){\bf \tau_1} {\bf \tau_2} ) \, 
{\bf \sigma_1} {\bf \sigma_2} \right]
\delta ({\bf r} - {\bf r'})
\label{F0}
\eeq
where the normalization factor
$C_0 = (dn/d\eps_{\Fs})^{-1}$ is the inverse density of
states at the Fermi surface.

In the FFS theory, a strong dependence of the scalar-isoscalar
amplitude $f_0$ on the observation point ${\bf r}$ was introduced.
In fact, in Ref.~\cite{AB1} a simple interpolation form of such a 
dependence was suggested:
\beq
f_0({\bf r}) =  f^{\exx} +   
(f^{\inn} - f^{\exx}) \frac {\rho_+({\bf r})} {\rho_0}.
\label{int}
\eeq
Here $\rho_+ ({\bf r})= \rho_n({\bf r}) + \rho_p({\bf r}) $ is the nuclear 
density in the point ${\bf r}$, while $\rho_0 = \rho_+(r{=}0)$. 
The subscript "0" for the zero Landau harmonics is for
brevity omitted in the r.h.s. of Eq.~(\ref{int}) and below.

It is worth to mention that the density dependence of
the phenomenological Skyrme force \cite{VB} agrees with the
ansatz of Eq.~(\ref{int}). There exist also alternative versions
of the interpolation formula for $f_0({\bf r})$ by means of 
the function $(\rho / \rho_0)^{\alpha}, {\alpha} \neq 1$, in the 
r.h.s.  of Eq.~(\ref{int})
or a more complicated density dependence
\cite{KhS, KhSZ, Fay2, Fay1}.
But all of them are characterized by a strong difference between
the dimensionless parameters $f^{\exx}$ and $f^{\inn}$: 
$f^{\exx} \simeq -3$, whereas  $f^{\inn}$ is close to zero.
\footnote{The inequality $f^{\inn} > -0.5$ should be fulfilled
to avoid the Pomeranchuck instability \cite{AB1} } 

It should be noted that the density dependent scalar-isoscalar
amplitude
\beq
f'_0({\bf r}) =  f'^{\exx} +   
(f'^{\inn} - f'^{\exx} ) \frac {\rho_+({\bf r})} {\rho_0}
\label{int1}
\eeq
was used in a new 
version of the energy functional method by S.Fayans 
et al.~\cite{Fay1,Fay2}
based on a detailed analysis of long isotopic chains.
The difference
between the parameters $f'^{\inn}, f'^{\exx} $ is also significant,
though not so dramatic as in the isoscalar case.
In principle, relations similar to Eq.~(\ref{int}) can be written
also for the spin-dependent terms of Eq.~(\ref{F0}), but up to now
no evidence of a noticeable difference between the internal
and external values of these amplitudes was found. Therefore the
equalities $g^{\inn} = g^{\exx}$, $g'^{\inn} = g'^{\exx}$
were imposed in the FFS theory.

In Ref.~\cite{BLSZ1} it was found that the external values of the
LM amplitudes can be calculated in terms of the
off-shell $T$-matrix of free $NN$-scattering taken at a negative energy
$E = 2\mu$, where $\mu$ is the chemical potential of the
nucleus under consideration. Stable nuclei with 
$\mu_n = \mu_p \simeq -8$\,MeV were considered, and a reasonable
agreement with the phenomenological values for the surface parameters
of the LM amplitude was obtained.
\footnote{They depend a little on the type of the density dependence.
The best agreement was achieved with the spin independent parameters
of Ref.~\cite{Fay1} and the spin dependent parameters of Ref.~\cite{Fay3}}
It confirmed the relevance of the asymptotic relation
$F \to T(2\mu)$ for the description of the properties of stable
nuclei.

For the spin independent amplitudes under consideration the
explicit form of these relations is as follows:
\beq
f_0^{\exx} = \frac {3}{16} \left[ t_0(E=2\mu) + t_1(E=2\mu)\right] ,
\label{fex0}
\eeq
\beq
f'^{\exx}_0= 
\frac {1}{16} \left[ t_0(E=2\mu)  - 3 t_1(E=2\mu)\right]  ,
\label{fex1}
\eeq
where  $t_0,t_1$ are the dimensionless values of the off-shell
$T$-matrix with the spin value $S=0$ and $S=1$ respectively,
taken at the zero value of all nucleon momenta.
In stable nuclei the isospin symmetry works well and one has:
\beq
f_{nn}^{\exx} = f_{pp}^{\exx} = f_0^{\exx} + f'^{\exx}_0= 
\frac {1}{4} t_0(E=2\mu) ,
\label{fexnn}
\eeq
\beq
f_{np}^{\exx} = f_0^{\exx} - f'^{\exx}_0= 
\frac {1}{8} \left[ t_0(E=2\mu)  + 3 t_1(E=2\mu)\right]  ,
\label{fexnp}
\eeq

In this paper these relations are applied to nuclei close to
the drip line in which the neutron and proton chemical potentials
deviate significantly from each other. As far as both 
amplitudes $t_0$ and
$t_1$ at small energy $E$ depend on $E$  rather sharply, the
isospin symmetry is destroyed. As a result, one 
deals with the situation when 
$f_{nn}^{\exx} \neq f_{pp}^{\exx}$. Both of these amplitudes can be
found from Eq.~(\ref{fexnn}), but at $E=2\mu_n$ for the first one
and at $E=2\mu_p$, for the second one. The amplitude 
$f_{np}^{\exx}$ can be obtained from Eq.~(\ref{fexnp}) at  
$E=\mu_n + \mu_p $. In the neutron drip line vicinity,
only the neutron chemical potential is close to zero,
whereas, on the contrary, $|\mu_p|$ increases with approaching 
the boundary. Therefore only the amplitude     
$f_{nn}^{\exx}$ changes significantly (see Fig.~1).

The SC relation, Eq.~(\ref{SCR}), with the amplitude $F$ given by
Eqs.~(\ref{F0}), (\ref{int}), (\ref{int1}) can be readily
integrated over for a spherical nucleus
yielding the following relation for the neutron potential:  
\bea
U_n(r) = C_0 \left[ 
f^{\exx}_{nn}\; \rho_n(r)  + (f^{\inn}_{nn} - f^{\exx}_{nn}) 
\frac {\rho_n(r) } {2\rho_0} \left( \rho_+(r) +\rho_p(r)\right) \right. 
\nonumber \\
\left. + f^{\exx}_{np} \; \rho_p(r)  + (f^{\inn}_{np} - f^{\exx}_{np}) 
\frac {\rho_p(r) } {2\rho_0} \left( \rho_+(r) + \rho_p(r) \right)\right].
\label{Un}
\eea
A similar relation is obtained for the proton potential $U_p(r)$ 
replacing the indices "$n$" and "$p$".

To estimate the effects of the energy (or $\mu$-) dependence
of the interaction amplitude $f^{\exx}_{nn}$, let us 
find the value of $U_n(0)$. To simplify the expression,
we use the approximations $\rho_n(0) = (N/A) \rho_0 $
and $\rho_p(0)= (Z/A) \rho_0$. Then we get:
\beq
U_n(0) = \frac{1}{2} C_0 \rho_0 \left[ 
f^{\exx}_{nn} \; \frac {N^2}{A^2} + 
f^{\inn}_{nn}  \left( 1 -\frac {Z^2}{A^2} \right) +
f^{\exx}_{np} \; \frac {Z^2}{A^2} + 
f^{\inn}_{np} \left( 1 -\frac {N^2}{A^2} \right)
\right].
\label{U0}
\eeq

Let us add  a small number of neutrons to a heavy nucleus. 
Then, neglecting for a while the pairing effects,
we obtain an approximate relation
\beq
\delta \mu_n  = \delta U_n(0)
\label{dmu}
\eeq
for the change of the neutron chemical 
potential.~\footnote{This equation contains also 
an additional term
provided the neutrons occupy a new $j$-level. All the
consideration presented below remains valid in this case, too.} 
The variation of the expression (\ref{U0}) is a sum of
two terms,
\beq
\delta U_n(0) = \delta U_1 + \delta U_2 \ ,
\label{dU12}
\eeq
having different origin. The first term results from variation
of $N$ and $A$ while the second one,
\beq
\delta U_2 = \frac{1}{2} C_0 \rho_0
\frac {N^2}{A^2} \delta f^{\exx}_{nn} \ ,
\label{dU2}
\eeq
is due to the $\mu$-dependence of the amplitude $f^{\exx}_{nn}$.
Neglecting this $\mu$-dependence one gets the accustomed variation
of the chemical potential
\beq
\delta \mu_n^{0}  = \delta U_1 .
\label{dmu0}
\eeq
But taking of the second term into account yields a dramatic
deviation from this traditional result
when nuclei with a big neutron excess are considered.

Indeed, at small values of $\mu_n$, the amplitude $f^{\exx}_{nn}$,
Eq.~(\ref{fexnn}), taken at $E=2\mu_n$ is singular:
\beq
f^{\exx}_{nn} = \frac {\alpha}{\sqrt{E}}= \frac {\alpha}{\sqrt{2\mu_n}}.
\label{pole}
\eeq
Here we neglected the small, compared to $\mu_n$,
value of the virtual level energy of the $T$-matrix
in the singlet channel. The variation of Eq.~(\ref{pole}) yields:
\beq
\delta f^{\exx}_{nn} = - f^{\exx}_{nn} \frac {\delta \mu_n }{2 \mu_n }.
\label{dfex}
\eeq
Upon substituting Eqs.~(\ref{dU12}), (\ref{dU2}), (\ref{dmu0}), 
and (\ref{dfex}) into Eq.~(\ref{dmu}) one finds
\beq
\delta \mu_n  = \frac {\delta \mu_n^0} {1 + V_0 /(2 \mu_n)} \ ,
\label{esdmu}
\eeq
where a short notation
$V_0 = \frac{1}{2} C_0 \rho_0 \frac {N^2}{A^2} f^{\exx}_{nn}$
for the first term of Eq.~(\ref{U0}) is introduced.

It can be easily seen that the denominator of the relation
(\ref{esdmu})
is noticeably greater than unit for nuclei with small value of $\mu_n$.
Let us, for example, calculate this quantity for two isotopes of tin
in the vicinity of the old drip line ($A_{max}=176$ which is the 
common value to all calculations for the tin isotopes as far as we 
know). We use the values of $\mu_n$
which will be found below and the corresponding values of the
amplitude $f^{\exx}_{nn}$ taken from Fig.1. We take also the 
standard values of the normalization parameters: $C_0=300\;$
MeV$\cdot$fm$^3$ and   $\rho_0=0.16\;$fm$^{-3}$.
First, let us consider the $^{150}$Sn nucleus. In this case, we have
$f^{\exx}_{nn}=-1.4$, $V_0=-15\;$MeV, and $\mu_n=-3.4\;$MeV.
The substitution of these values into Eq.~(\ref{esdmu}) yields
$\delta \mu_n = \delta \mu_n^0 /3.2 $.
The analogous calculation for the $^{200}$Sn isotope
($\mu_n=-2.0\;$MeV, $f^{\exx}_{nn}=-1.66$,
        $V_0=-22.4\;$MeV)
results in $\delta \mu_n = \delta \mu_n^0 /6.5$.
We see that deviations from the traditional scheme are
really large and are growing as soon as with the value of 
$|\mu_n|$ becomes less and less. This explains qualitatively
why  nuclei (e.g. $^{200}$Sn )
which are strongly unbound in traditional calculations
could exist in our approach.

Going to the actual calculations incorporating the
effect discussed above, we start from an advanced version
of the SC FFS theory, the so-called quasiparticle Lagrange method 
(QLM).~\footnote{The term "quasiparticle" is used here in Landau's 
(not Bogolyubov's) sense.}
It was devised in Ref.~\cite{KhS} 
for magic nuclei and extended in Refs.~\cite{ZS1,ZS2}
to include superfluidity. All necessary modifications in the
scheme can be explained for a simpler, nonsuperfluid case.
We present here only a brief sketch of the QLM referring to
Refs.~\cite{KhS,ZS1,ZS2} for details. The approach utilizes
the Lagrange formalism which is the most convenient 
when the energy dependence effects are considered
explicitly. The effective Lagrangian is constructed in such
a way that its variation results in the Dyson equation, 
\beq
(\eps - \eps^0_p -\Sigma_q)\, G_q = 1,  
\label{Dys}
\eeq
for
the quasiparticle Green function $G_q$ with the
quasiparticle mass operator
\beq
\Sigma_q({\bf r},{\bf k}, \eps) = 
\Sigma_0({\bf r}) +
{1 \over {(k^0_{\Fs})^2} }\, {\bf k}\, \Sigma_1({\bf r})\, {\bf k} +
{\eps \over {\eps^0_{\Fs}} }\, \Sigma_2({\bf r}),
\label{Sigma}
\eeq
where the normalization parameters are:
$k^0_{\Fs} = \pi /(m C_0) $,
$\eps^0_{\Fs} = (k^0_{\Fs})^2/(2m)$.

The first two terms of $\Sigma_q$ are common to the HF theory
with effective velocity dependent forces (e.g., Skyrme forces).
The third term, $\Sigma_2(r)$, is due to energy dependence. 
It determines the coordinate dependent $Z$-factor: 
\beq
Z({\bf r}) = \left( 1 -
 \Sigma_2({\bf r}) / \eps^0_{\Fs} \right)^{-1}.
\label{Zfac}
\eeq

The solution of the Eq.~(\ref{Dys}) can be expressed in terms
of the eigenfunctions $\Psi_{\lambda}$ of the corresponding
homogeneous equation as:
\beq
G_q({\bf r_1},{\bf r_2},\eps) = \sum_{\lambda} {
\frac { \Psi^*_{\lambda}({\bf r_1}) \Psi_{\lambda}({\bf r_2}) }
{\eps - \eps_{\lambda} + i \delta {\rm sgn} (\eps_{\lambda} - \mu) } },
\label{G-pol}
\eeq
where $\delta $ is small and positive.
The functions $\Psi_{\lambda}$ are orthonormalyzed with the weight: 
\beq
\int d{\bf r}\, \Psi^*_{\lambda} ({\bf r}) Z^{-1} ({\bf r})
\Psi_{\lambda' }({\bf r})  = \delta_{\lambda \lambda' }.
\label{ort}
\eeq
   
The quasiparticle density $\nu_0 (\bf r)$ 
associated with the $\Psi_{\lambda}$-functions is:
\beq
\nu_0 ({\bf r}) = \sum_{\lambda} 
n_{\lambda} |\Psi_{\lambda}({\bf r})|^2,
\label{nu}
\eeq
where $n_{\lambda} = (0,\; 1)$ are the quasiparticle occupation numbers.
It differs from the usual density $\rho (\bf r) $
normalized to the total particle number by the $Z$-factor: 
\beq
\nu_0 ({\bf r}) = Z ({\bf r})\, \rho ({\bf r}).
\label{nunorm}
\eeq

There are two additional densities introduced in
Ref.~\cite{KhS}, the quasiparticle kinetic
energy density 
\beq
\nu_1 ({\bf r}) = {1 \over {(k^0_{\Fs})^2} }\sum_{\lambda} 
n_{\lambda} |\nabla \Psi_{\lambda}({\bf r})|^2,
\label{nu1}
\eeq
and the total quasiparticle energy density
\beq
\nu_2 ({\bf r}) = {1 \over {\eps^0_{\Fs}} }\sum_{\lambda} 
n_{\lambda} \eps_{\lambda} |\Psi_{\lambda}({\bf r})|^2,
\label{nu2}
\eeq

The density $\nu_1({\bf r})$ is analogous to the quantity
$\tau({\bf r})$ of the HF theory \cite{VB}, whereas the
density $\nu_2({\bf r})$ is a new ingredient of the QLM,
which does not appear in the HF approach.

In notation of Ref.~\cite{KhS}, the density of the interaction 
Lagrangian $L'$ is
\bea
{\cal L}' ({\bf r}) = - C_0 \left[ 
 \frac {\lambda_{00}} {2} \nu_{0+}^2({\bf r}) + 
 \frac {\lambda'_{00}} {2} \nu_{0-}^2({\bf r}) + 
 \frac {2 \gamma} {3 \rho_0^0} \nu_{0+}({\bf r}) 
 \nu_{0n}({\bf r})  \nu_{0p}({\bf r})+
 \lambda_{01} \nu_{0+}({\bf r}) \nu_{1+}({\bf r}) \right. 
\nonumber \\
\left. + \lambda'_{01} \nu_{0-}({\bf r}) \nu_{1-}({\bf r}) + 
 \lambda_{02} \nu_{0+}({\bf r}) \nu_{2+}({\bf r}) -
 \frac {\lambda_{00} r_0^2} {2} ( \nabla \nu_{0+}({\bf r}))^2 
\right] + {\cal L}_1,   
\label{Lint}
\eea
where the normalization density is $\rho_0^0 = 2 (k^0_{\Fs})^3/(3\pi^2)$,
and ${\cal L}_1$ includes the Coulomb and the spin-dependent terms
(mainly, the spin-orbit one). It is worth to mention that the isotopic
structure of the "triple" term in Eq.~(\ref{Lint}) (proportional to 
$\gamma$) is similar to that of the Skyrme Hamiltonian of Ref.~\cite{VB}.

To extract from the Lagrangian $L'$ the LM amplitude $F$, 
which, in accordance with the Landau prescription, is the second 
derivative of the ground state energy,
one needs to find the 
corresponding Hamiltonian expressed in terms of the usual
densities $\rho(\bf r)$ and $\tau({\bf r})$. Then it is necessary
to calculate the second
variational derivative with respect to $\rho$. This leads
to a rather cumbersome expression which can be found
in Ref.~\cite{BLSZ2}. However, the relation of
the constants $\lambda_{00} , \lambda'_{00} $ with the external LM
amplitudes $f^{\exx}, f'^{\exx}$ in Eqs.~(\ref{int}), (\ref{int1})
can be understood without using  any explicit formula.
Indeed, in the asymptotic region
outside the nucleus the $Z$-factor is tending to unit and
the densities $\nu_0(\bf r)$ and $\rho(\bf r)$ coincide.
Therefore the variation of $L'$ with respect to $\rho$
can be replaced by the variation with respect to $\nu_0$.
The amplitudes under consideration originate from the first
three terms of Eq.~(\ref{Lint}). But the second variation
of the third term vanishes at large $r$ and one ends up with 
the following identities:
\beq
\lambda_{00} = f^{\exx}, \quad \lambda'_{00} = f'^{\exx}. 
\label{lam0_01}
\eeq
 Of course, they follow also from the explicit relation
for $F$ of Ref.~\cite{BLSZ2}.

Dealing with superfluid nuclei, we utilize the modification of 
QLM developed for this case in Refs.~\cite{ZS1,ZS2}.
Although the method itself is, in principle, rather general,
the practical scheme of Refs.~\cite{ZS1,ZS2} is quite
simple. The main approximations 
are as follows. First, the $\lambda$-representation 
with a limited $\lambda$-basis is used                    
($\eps_{\mbox{\scriptsize min}} <
\eps_{\lambda}  < \eps_{\mbox{\scriptsize max}}, \; 
\eps_{\mbox{\scriptsize min}} = - (20 \div 25)\;$MeV,
$\eps_{\mbox{\scriptsize max}} =5\;$MeV), with the discretization 
of the continuum. 
Second, the $\delta$-form density independent ("volume")
pairing interaction is considered with
the strength depending on the basis, in accordance with the prescription of
Ref.~\cite{AB1}: $\Gamma_{\xi} = C_0 \ln^{-1}(C_p/\xi)$,
where $\xi = \sqrt{(\mu - 
\eps_{\mbox{\scriptsize min}} )(\eps_{\mbox{\scriptsize max}} - \mu)}$.
At last, the diagonal approximation for the gap $\Delta$
is used: $\Delta_{\lambda \lambda'} = \Delta_{\lambda} 
\delta_{\lambda \lambda'} $.
Thus, the method of describing the pairing effects 
in Refs.~\cite{ZS1,ZS2} in the main points coincides with the BCS
approximation. Although such a scheme possesses some well-known deficiencies
for nuclei near the drip line and there exist  much more advanced
approaches for this case (see, e.g., Refs.~\cite{Dob2,Fay1}), we conserve 
here all details of 
the pairing scheme \cite{ZS1,ZS2} (including the values of the 
pairing parameters)
in order to separate the effect of the energy
dependence of the mean field more clearly.

Our goal is to single out effects of the energy dependence
of the external values of the LM amplitude. For convenience, let us
rewrite the first three terms of Eq.~(\ref{Lint})
substituting the quantities
$f^{\exx},f'^{\exx}$ for $\lambda_{00} ,\lambda'_{00} $: 
\beq
{\cal L}_0 ({\bf r}) = - C_0 \left[ 
 \frac {f^{\exx}} {2}\, \nu_{0+}^2({\bf r}) + 
 \frac {f'^{\exx}} {2}\, \nu_{0-}^2({\bf r}) + 
 \frac {2 \gamma} {3 \rho_0^0}\, \nu_{0+}({\bf r})', 
  \nu_{0n}({\bf r})\, \nu_{0p}({\bf r}) \right]. 
\label{L0}
\eeq

If we consider $f^{\exx},f'^{\exx}$ as phenomenological parameters,
there is no difference between this expression and the initial one,
just the physical meaning of the constants $\lambda_{00},\lambda'_{00}$  
is more transparent. The next step is the calculation
of these parameters from Eqs.~(\ref{fex0}), (\ref{fex1}). They are 
$f^{\exx}= -2.6$ and $f'^{\exx}=1.56$, Ref.~\cite{BLSZ1}, instead of
$\lambda_{00} = -3.25$ and $\lambda'_{00} = 2.4$ of Ref.~\cite{KhS}
\footnote{These values of $f^{\exx},f'^{\exx}$ are
closer to those of Ref.~\cite{Fay1}.}. 
Now Eq.~(\ref{L0}) contains only one adjustable parameter $\gamma$
instead of the three those of the corresponding part
of Eq.~(\ref{Lint}).
We chose it in such a way to better reproduce the single-particle energy 
spectrum of the nucleus $^{124}$Sn calculated in Ref.~\cite{ZS1}
where it was used for fitting the parameters of the calculation scheme
and was found in a reasonable agreement with the experimental one.
It should be noted that we do not analyze here the total binding
energies and the density distributions but
concentrate on the  single-particle spectra because they are mainly
responsible for the position of the drip line. We found that the spectrum
obtained with the {\it ab initio} constants $f^{\exx},f'^{\exx}$  
and with the value of $\gamma = 1.6$ (instead of $\gamma = 3.2$ in 
Refs.~\cite{KhS,ZS1})
is in a reasonable agreement with that of Ref.~\cite{ZS1} (see Table 1),
where $E_{\lambda} = \mu \pm \sqrt{(\eps_{\lambda} -\mu)^2 +
\Delta_{\lambda}^2}$.

The generalization of Eq.~(\ref{L0}) to nuclei with a large neutron 
excess, where the isotopic symmetry is violated, 
is quite obvious:
\bea
\tilde {\cal L}_0 ({\bf r}) = C_0 \left[ 
 \frac {1}{2}  f^{\exx}_{nn}(E=2\mu_n)\; \nu_{0n}^2({\bf r}) + 
 \frac {1}{2}  {f^{\exx}_{pp}(E=2\mu_p)}\; \nu_{0p}^2({\bf r})  \right. 
\nonumber \\
\left. +  f^{\exx}_{np}(E=\mu_n + \mu_p) \; \nu_{0n}({\bf r}) 
\nu_{0p}({\bf r})+ 
\frac  {2 \gamma} {3 \rho_0^0}\, \nu_{0+}({\bf r})\,  
\nu_{0n}({\bf r})\, \nu_{0p}({\bf r})  \right] ,  
\label{L0nn}
\eea
where the external interaction LM amplitudes 
$f^{\exx}_{nn}, f^{\exx}_{pp}, f^{\exx}_{np} $
are energy dependent and should be calculated for the nucleus
under consideration in the same way as in Eq.~(\ref{Un}).
Again the energy dependence of the first of them only is essential
in the case under consideration. In the two next amplitudes, 
this dependence is retained just for presenting a more
general form which is essential, e.g., for nuclei near the
proton drip line.

We made a series of the self-consistent  calculations 
for the chain of the tin isotopes, first,  with the
phenomenological Lagrangian, 
Eq.~(\ref{Lint}),~\footnote{They just repeat the calculations 
of Ref.~\cite{ZS2}}
 and, second, with the semi-microscopic one,
Eq.~(\ref{L0nn}). Results are displayed in Fig.~2, together with the 
predictions of Dobaczewski et al. \cite{Dob2} and Fayans et al. 
\cite{Fay1}. The latter two approaches, just as the one of 
Ref.~\cite{ZS2},  use the phenomenological energy-independent
forces \footnote{The trivial, linear energy dependence, 
taken into account in Ref.~\cite{ZS2},  which is incorporated
into Eq.~(\ref{Lint}) via the density $\nu_2({\bf r})$,
is not important for the effect under consideration}.
The results of the three phenomenological calculations are
quite close to each other. Qualitatively, they could be considered
as one "phenomenological" curve which is in very good agreement
with available experimental data. At small and intermediate values
of the asymmetry parameter $y=(N-Z)/A$ the results of our 
semi-microscopic calculation deviate
from the phenomenological curve (and, consequently, from the
experimental data), but not significantly. The reason of this
deficiency is quite simple. The matter is that the one-parameter
Eq.~(\ref{L0nn}) does not permit to obtain simultaneously reasonable
values of the two "inner" amplitudes, $f^{\inn}$ and $f'^{\inn}$.
It is well known that the Skyrme prescription of the isotopic structure
of the "triple" term used in Eq.~(\ref{L0nn}) is not obligatory for
the effective force. The simplest two-parameter generalization is:
\beq
\frac  {2 \gamma} {3 \rho_0^0}\, 
\nu_{0+}({\bf r}) \,  
\nu_{0n}({\bf r}) \,\nu_{0p}({\bf r})  \to 
\frac  {1 } {6 \rho_0^0} \left[ 
\gamma  \nu^3_{0+}({\bf r}) + 
\gamma'  \nu_{0+}({\bf r}) \nu^2_{0-}({\bf r}) \right],  
\label{L0nn1}
\eeq
which reduces to Eq.~(\ref{L0nn}) at $\gamma' = -\gamma$.
The ansatz of Eq.~(\ref{L0nn1}) seems to be more adequate to
ideas underlying the approach suggested. Such a generalization,
with a new adjustment of the parameters, will be carried out
in a separate work. Preliminary estimates show that in this case the 
"semi-microscopic" results for stable nuclei become much closer
to the phenomenological ones. 

When we approach to the drip line, deviations grow,
the semi-microscopic curve being significantly higher than
the phenomenological one. 
It is remarkable that our calculation predicts the existence 
of nuclei beyond the commonly admitted end
of the tin chain, $A_{\mbox{\scriptsize max}}=176$. 
It is worth to mention that this value of $A_{\mbox{\scriptsize max}}$
corresponds to the value of $y=0.43$ which exceeds the critical value 
$y_0=0.37$ predicted for the asymmetric nuclear matter \cite{asym}. 
The energy dependence effects under
consideration make this difference between finite nuclei and 
nuclear matter more pronounced. We interrupted our calculations
at $A=208$ just because we are conscious that the deficiencies of
the approach become more serious with the value of $|\mu_n|$ 
becoming less. One of them is too schematic consideration
of pairing. Then, corrections to the standard FFS theory could appear
when the energy dependence of the LM amplitude is significant. 

It is worth to note that the two recipes of taking into
account the energy dependence effects, Eq.~(\ref{Un}) and
Eq.~(\ref{L0nn}), are not identical. Indeed, in the first one
the amplitudes $f^{\inn}, f'^{\inn}$ are completely energy
independent. As to the second recipe, the parameter 
$\gamma$ is supposed to be energy independent that leads
to some energy dependence of the 
quantities $f^{\inn}, f'^{\inn}$
evaluated via the procedure described above.
In the framework of a pure phenomenology,
it is difficult to choose between the two ansatz. 
In our regular calculations, we use the second choice just
because we have a well developed calculation scheme for this case.
A more consistent microscopic theory for solving this problem is 
necessary. Such an approach based on the Brueckner theory
for finite nuclei is now in progress \cite{BLSZ2,BLSZ3},
but it is yet far from to be completed.  
A semimicroscopic model for the scalar-isoscalar LM
amplitude $f({\bf r})$ suggested in Ref.~\cite{BLSZ2}
on the base of an approximate calculation
of the $G$-matrix for a slab of nuclear matter  \cite{BLSZ3}
shows some energy dependence of the parameter $f^{\inn}$,
but it is rather smooth. Evidently, the truth is somewhere
between the two ansatz discussed above. What is 
important for our preliminary analysis, that
both of them predict qualitatively the same. Namely, the energy
dependence of the external LM amplitude $f^{\exx}_{nn}$ 
makes the neutron potential well deeper 
when the absolute value of $\mu_n$ becomes less. This effect
could help nuclei with a huge neutron excess to survive.

The calculation scheme proposed in this letter possesses some 
evident deficiencies partially discussed above.
Improvement of them could make the effect less pronounced.
That is the reason why the title of the paper is formulated 
as a question. However, the effect itself is 
based on a transparent physical phenomenon.
Therefore we expect that it should remain, at least 
qualitatively, after all
necessary corrections to the approximate calculation
presented in this letter.

This research was partially supported by Grants
No.~00-15-96590 and No.~00-02-17319 from the Russian
Foundation for Basic Research.
We thank S.V.~Tolokonnikov for useful
discussions and for supplying us with the numerical
data used in Ref.~\cite{Fay1}.

{}

\newpage

\begin{center}
{\bf Table 1}

\vskip .5 cm
Single-particle spectrum of the $^{124}$Sn nucleus.

\vskip .5 cm

\begin{tabular}{|c|cc|ccc|}
\hline
\rule[-7pt]{0pt}{20pt} $\lambda$  &
   \multicolumn{2}{|c|}{$\varepsilon_{\lambda}$, Mev} &
   \multicolumn{3}{|c|}{$E_{\lambda}$, Mev} \\
\cline{2-6} &
\rule[-5pt]{0pt}{15pt} \cite{ZS1} & this work & \cite{ZS1}  & this work & 
exp. \\
\hline
\rule{0pt}{18pt}
$2p_{1/2}$  & $-16.86$ & $-17.94$ & $-16.96$ & $-18.05$ &         \\
$1g_{9/2}$  & $-15.15$ & $-15.93$ & $-15.29$ & $-16.11$ &         \\
$2d_{5/2}$  & $-10.19$ & $-10.82$ & $-10.42$ & $-11.05$ &         \\
$3s_{1/2}$  &  $-9.53$ &  $-8.82$ &  $-9.97$ &  $-9.18$ & $-8.64$ \\
$1g_{7/2}$  &  $-8.34$ &  $-8.32$ &  $-8.81$ &  $-9.19$ & $-9.63$ \\
$2d_{3/2}$  &  $-8.08$ &  $-8.05$ &  $-8.68$ &  $-8.66$ & $-8.52$ \\
$1h_{11/2}$ &  $-7.19$ &  $-6.87$ &  $-5.83$ &  $-5.08$ & $-5.73$ \\
$2f_{7/2}$  &  $-2.34$ &  $-2.29$ &  $-2.24$ &  $-2.14$ &         \\
$3p_{3/2}$  &  $-0.97$ &  $-0.96$ &  $-0.94$ &  $-0.91$ &         \\
\rule[-10pt]{0pt}{10pt}
$3p_{1/2}$  &  $-0.42$ &  $-0.26$ &  $-0.39$ &  $-0.23$ &         \\
\hline
\end{tabular}

\end{center}

\newpage

\centerline{\bf Figure captions}
\begin{enumerate}

\item 
External LM amplitudes 
$f^{\exx}_{nn}(E)$, $f^{\exx}_{pp}(E)$, and $f^{\exx}_{np}(E)$
(the latter is divided by 10) taken with negative sign. The solid lines 
show these amplitudes within the energy limits which are typical for 
tin isotopes.

\item 
A half of two-neutron separation energy $S_{2n}/2$ calculated for even-even 
tin isotopes. Results of the semi-microscopic calculations of this work 
(solid line) 
are displayed together with predictions of Ref.~\cite{ZS2} (dashed 
line), Ref.~\cite{Dob2} (triangles) and Ref.~\cite{Fay1} (squares). 
Experimental values of $S_{2n}/2$ are shown by circles. 

\end{enumerate}

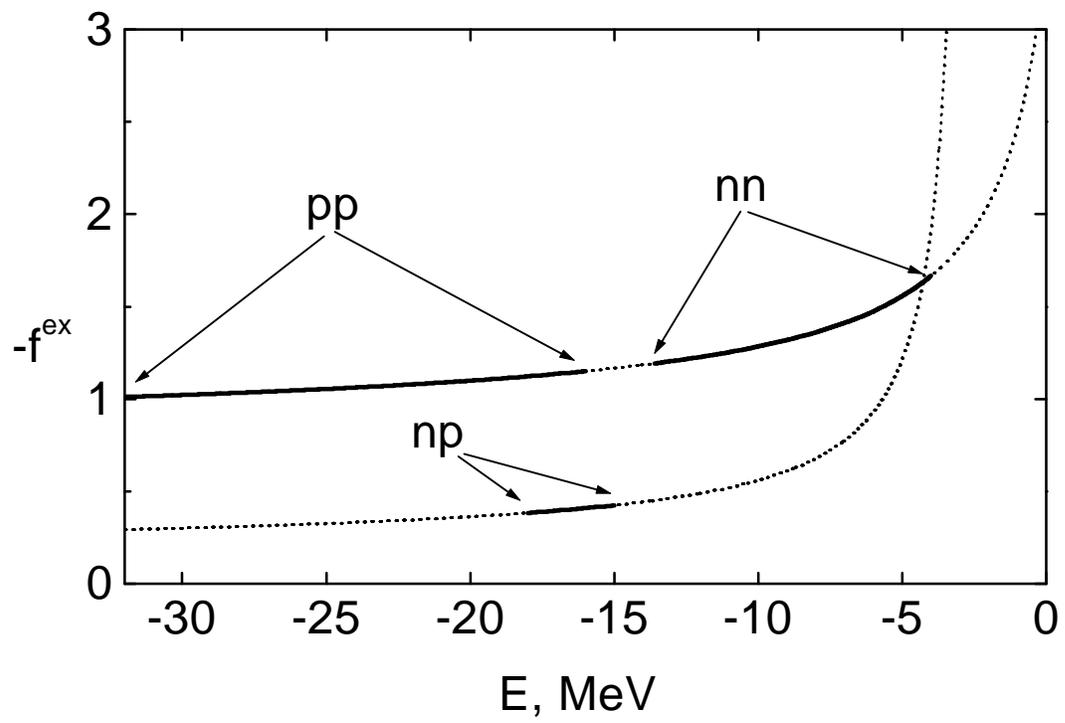
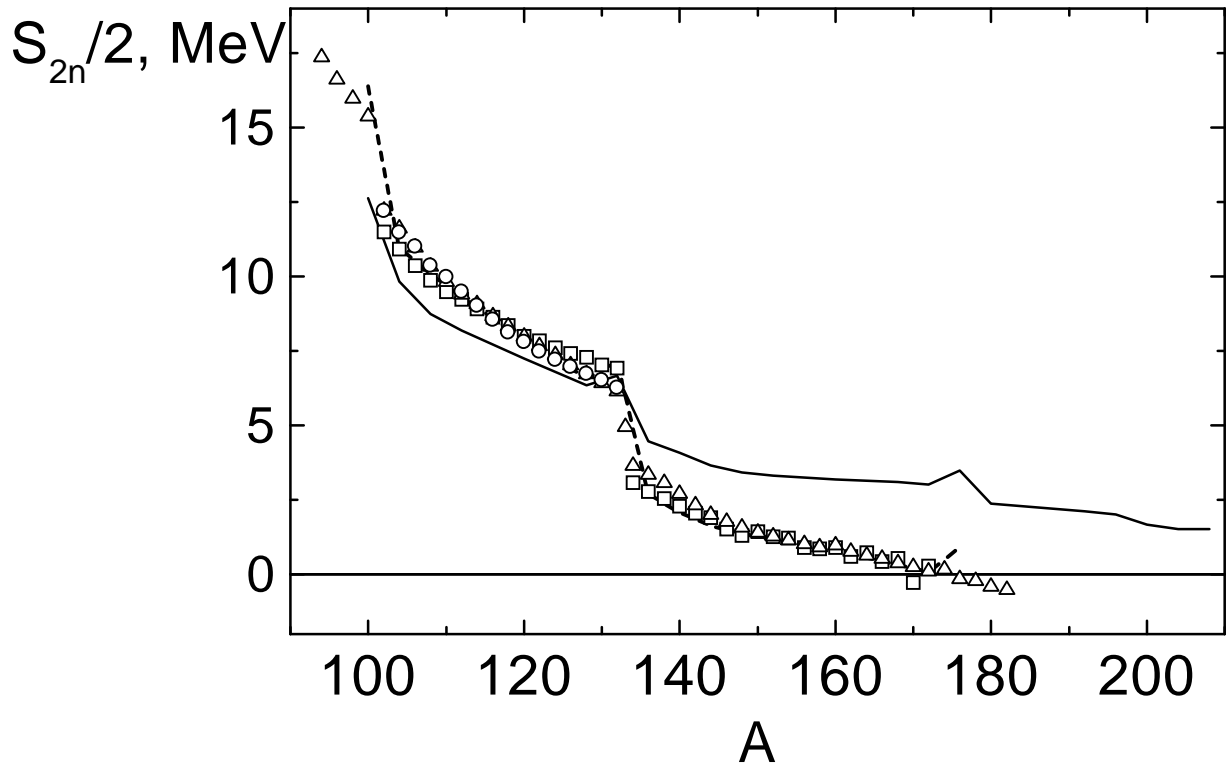
\end{document}